\documentclass[twocolumn,noshowpacs,preprintnumbers,amsmath,amssymb]{revtex4}

\usepackage{verbatim}

\usepackage{graphicx}
\usepackage{dcolumn}
\usepackage{bm}
\usepackage{wrapfig}

\begin{document}

\title{Preparation and electrical properties of cobalt-platinum nanoparticle monolayers deposited by the Langmuir-Blodgett technique}
\author{Vesna Aleksandrovic}
\author{Denis Greshnykh}
\author{Igor Randjelovic}
\author{Andreas Fr\"{o}msdorf}
\author{Andreas Kornowski}
\author{Stephan Volkher Roth}
\author{Christian Klinke}
\author{Horst Weller}
\email{weller@chemie.uni-hamburg.de}
\affiliation{Institute of Physical Chemistry, University of Hamburg, D - 20146 Hamburg, Germany}

\begin{abstract} 

The Langmuir-Blodgett technique was utilized and optimized to produce closed monolayers of cobalt-platinum nanoparticles over vast areas. It is shown that sample preparation, "dipping angle", and subphase type have a strong impact on the quality of the produced films. The amount of ligands on the nanoparticles surface must be minimized, the dipping angle must be around 105$^{\circ}$, while the glycol subphase is necessary to obtain nanoparticle monolayers. The achieved films were characterized by scanning electron microscopy (SEM) and grazing incidence x-ray scattering (GISAXS). The electrical properties of the deposited films were studied by direct current (DC) measurements showing a discrepancy to the variable range hopping transport from the granular metal model, and favoring the simple thermal activated charge transport. SEM, GISAXS as well as DC measurements confirm a narrow size distribution and high ordering of the deposited films.

\end{abstract}

\maketitle

The properties of organized nanoparticle assemblies have intrigued many groups reflected in numerous publications~\cite{C01,C02,C03,C04,C05,C06}. Possible applications include sensors~\cite{C07,C08}, optical~\cite{C09,C10}, and electronic devices~\cite{C11,C12,C13,C14}. Furthermore, applications with a necessity of tailored materials are discussed in literature~\cite{C15}. Ordered arrangements of nanoparticles can be classified by their dimensionality~\cite{C06}. One-dimensional nanoparticle chains, two-dimensional arrays and super crystals representing the three-dimensional case can be distinguished. Two-dimensional ordering can be achieved by spin coating, dip coating, and the Langmuir-Blodgett (LB) technique. Latter has been widely used to assemble amphiphilic molecules. Combined with microcontact printing the LB technique has been utilized previously to assemble structured monolayers of nanoparticles in two-dimensional Au~\cite{C16}, Pt/Fe2O3~\cite{C17}, and Co~\cite{C18} patterns. The standard LB technique implies the film preparation by deposition of the particles onto water. Depending on the particle functionalization different approaches can be chosen. Particles surrounded by a hydrophobic ligand shell can be deposited onto water surfaces~\cite{C14,C19,C20,C21,C22}. In the case of low particle stability on the interface organic molecules can be utilized as Langmuir film which helps to deposit the particles onto water surfaces~\cite{C23,C24}. In the here presented approach we deposit, to our knowledge for the first time, nanoparticles onto the glycol/air interface. This yields highly ordered cobalt-platinum nanoparticle films over unprecedented vast areas in the micrometer up to the millimeter range.

The electrical properties of nanoparticle arrangements met a special interest in the scientific community~\cite{C11,C12}. Aside from other interesting effects like single electron tunneling~\cite{C13}, environment~\cite{C07}, and pressure~\cite{C25} sensitive conductivity, tunable properties~\cite{C04} of the nanoparticle arrays make them an interesting model system for charge transport studies in confined systems~\cite{C26,C27}. The here presented LB technique was used to deposit the particles onto samples structured with gold electrodes for DC measurements. In the temperature range between 80~K and 300~K they revealed that the film follows a simple thermally activated charge transport model with an activation energy of 18~meV.

\section*{Results and discussion}

\subsubsection*{Preparation of cobalt-platinum nanoparticle films using Langmuir-Blodgett technique}

Cobalt-platinum nanoparticles synthesized according to Ref.~\cite{C28} are stabilized by two types of ligands: adamantane carboxylic acid (ACA) and hexadecylamine (HDA). These ligands make the nanoparticles surface hydrophobic and therefore soluble in organic solvents such as chloroform or toluene. The hydrophobicity of the nanoparticles surface opens the possibility of nanoparticle monolayer film preparation by means of the Langmuir-Blodgett (LB) technique. However, the preparation of LB films in the classical way, i.e. using water as subphase results in films with poor substrate coverage. All attempts to obtain films of ordered particles on the water surface failed due to nanoparticle agglomeration and their later sinking caused by their comparatively high density. A possible reason for such behavior might be the ability of HDA to form Langmuir films on the water surface. In this way HDA occupies the interface, which in turn results in a poor surface coverage by nanoparticles. Additionally spreading of HDA on the water/air interface might facilitate its detachment from the nanoparticle surface and consequently deteriorate the nanoparticle stability. This may explain the observed agglomeration of nanoparticles on the water surface. 

In order to overcome the low surface coverage, as well as nanoparticle agglomeration another subphase were introduced. We chose ethylene glycol (EG) and diethylen glycol (DEG) as most common polar solvents with low evaporation rates in which the investigated nanoparticles are not soluble. In order to compare the behavior of HDA ligands on water and DEG subphases 110~$\mu$l of HDA solution in chloroform (c = 0.08 mol/l) was spread on the water/air and DEG/air interface. The compression of the obtained film was performed after chloroform evaporation, approximately 10-15 min later. The surface pressure ($\pi$) versus area (A) isotherm (Figure 1a) for HDA on water shows three phases characteristic for a LB film compression: the so-called "gas phase" below 120~cm$^{2}$, the "liquid phase" between 30 and 120~cm$^{2}$ and at the end the "solid phase" below 30~cm$^{2}$ of trough area. The same amount of HDA spread on DEG did not show any influence on the surface pressure, indicating that no LB-film was formed on this interface. Therefore, the complete EG or DEG interface is available for the nanoparticles, which was not the case on the water surface where HDA occupied large areas. The next step was to investigate whether cobalt-platinum nanoparticles can form Langmuir films on EG and/or DEG surfaces. 20~$\mu$m of 0.01 mol/dm$^{3}$ spherical nanoparticles solution (diameter $\cong$ 8.0 nm; see supp. inform.) were spread on both EG and DEG subphases. Two distinct isotherms recorded during the compression of these films are shown in Figure 1b. 

\begin{figure}[!h]
\begin{center}
\includegraphics[width=0.45\textwidth]{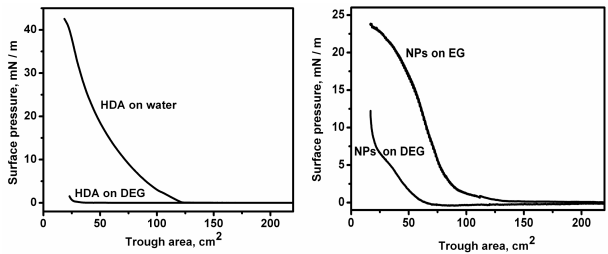}
\caption{\it Surface-pressure isotherms obtained by spreading: a) HDA on water and DEG; b) cobalt-platinum nanoparticles (NPs) on EG and DEG.}
\label{F01}
\end{center}
\end{figure}

By closing the barrier of the LB trough i.e. decreasing the area available for nanoparticles the surface pressure on EG remains almost constant up to approx. 130~cm$^{2}$. During this period, individual nanoparticles do not form continuous networks what is characteristic for the "gas phase" in LB films. Then, at smaller values of surface area, the surface pressure starts to increase indicating the beginning of the so-called "liquid phase" and the formation of nanoparticle networks. The surface pressure increases gradually up to a value of 80~cm$^{2}$ when the slope becomes steeper indicating even closer nanoparticle packing and formation of large domains of organized nanoparticles. Decreasing the surface area to the minimal value of 18~cm$^{2}$ the so-called "solid-state" region, characterized by a very steep isotherm and most densely packed nanoparticles was not achieved. 

In contrary, spreading the nanoparticles on a DEG subphase the surface pressure starts to increase at around 70~cm$^{2}$. At approximately 30~cm$^{2}$ the slope of the isotherm becomes very steep and close to linear what is typical for the "solid-state" phase. In this phase, the densest packing of the nanoparticles is achieved. Since this is not possible for EG surfaces, for all further investigations DEG was selected as subphase for film preparations. 

Apart from suppression of the HDA film formation, an additional reason for better spreading of cobalt-platinum nanoparticles on glycols subphase compared to water might be the solvents polarity. Namely, due to their stabilization by means of organic ligands, nanoparticles are hydrophobic, what makes their spreading on polar water surface very difficult. However, the considerably less polar EG and DEG surface allows nanoparticles film formation, which was also experimentally observed. More insight into the polarity of the solvent can be obtained from the static relative dielectric constant ($\epsilon_{r}$). For water $\epsilon_{r}$ is 80.10, while for EG it is 41.40 and for DEG 31.82. Latter value is almost two times lower than the one for water. Furthermore, the lower polarity of DEG compared to EG might be the reason for better spreading of cobalt-platinum nanoparticles on the DEG subphase.

Based on area-surface pressure isotherms recorded on the DEG subphase for cobalt-platinum nanoparticle solutions, the pressure of 8~mN/m was selected as an optimum for the film deposition on a wafer (Figure 1b). At this surface pressure, nanoparticles are packed in the solid-state phase. A further decrease of the surface area would increase the probability for particle overlapping, i.e. the formation of double layers.

In order to achieve a close packing of nanoparticles a barrier speed of 5~mm/min was selected, which allows the relaxation of the nanoparticle film. Simultaneously, a program called "Isothermal cycles" was performed in order to improve particles ordering (see experimental part). Finally, the film was deposited on a silicon wafer surface at a surface pressure of 8~mN/m.

\begin{figure}[!h]
\begin{center}
\includegraphics[width=0.45\textwidth]{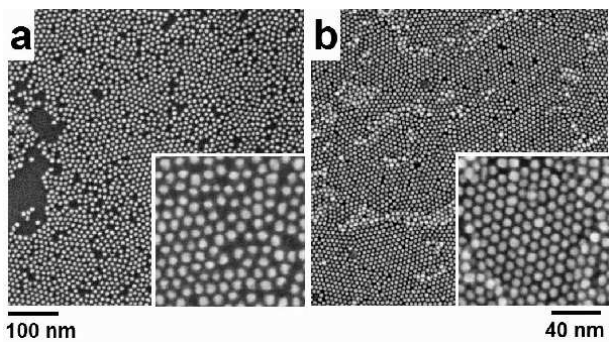}
\caption{\it SEM images of samples prepared from cobalt-platinum nanoparticle (d = 8.0~nm) solution (a) directly after syntheses (stock solution) and after additional washing procedures (b). Both samples were deposited from a DEG/air interface onto a silicon wafer at a surface pressure of 8-10~mN/m, respectively and at a substrate/surface angle of 105$^{\circ}$.}
\label{F02}
\end{center}
\end{figure}

Scanning electron microscopy (SEM) images of a nanoparticle film deposited onto a silicon wafer are shown in Figure 2a. Well-ordered domains of nanoparticles are visible as bright areas in the SEM image, while dark areas are correlating to domains not covered with nanoparticles. In order to improve the particle packing they were additionally washed twice (see experimental part) and deposited afterwards using the same program. In contrast to the initial particle solution, the additionally washed particles yield fully covered films as shown in the SEM micrograph in Figure 2b. As mentioned before, amphiphilic HDA molecules can form large aggregates on the subphase. The experiments show that in order to avoid non-continuous nanoparticle films the ligand excess present in the initial solution should to be removed by washing procedures. Nevertheless, both films consist of particles packed in well-ordered domains with different orientation. The double layers of nanoparticles were formed at domain boundaries in areas where these domains overlap (brighter areas in the SEM image, Fig. 2b). These investigations demonstrate that removing ligand excess from the initial cobalt-platinum nanoparticle solution is necessary to obtain dense and full substrate coverage.

\subsubsection*{Influence of the substrate/subphase angle on the film preparation}

Our investigations show that the dipping angle has a pronounced influence on the substrate coverage. The angle was varied from 105$^{\circ}$ (angle $\alpha$, Figure 3) to 180$^{\circ}$ (the wafer is normal to the subphase surface). All nanoparticle films were prepared by pulling the silicon substrate out of the subphase at different angles. The SEM images of the resulting films are shown in Figure 3.

\begin{figure}[!h]
\begin{center}
\includegraphics[width=0.45\textwidth]{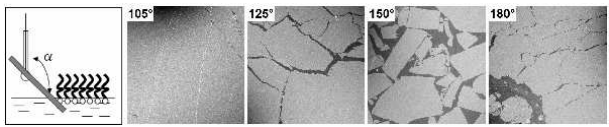}
\caption{\it SEM images: Overview of the cobalt-platinum nanoparticle (d = 8.0 nm) films deposited under different angles   onto silicon wafers. The bright areas correspond to particles and the dark areas correspond to the substrate.}
\label{F03}
\end{center}
\end{figure}

All samples taken under angles larger than 105$^{\circ}$ showed ruptures in the obtained films. At 125$^{\circ}$ and 180$^{\circ}$ the films exhibited long crevices, while at 150$^{\circ}$ individual smaller sheets were recognizable in the SEM image. The best results, considering substrate coverage, were obtained for the wafer fixed under 105$^{\circ}$. Hence, the following investigations were performed on films coated under this angle from DEG as subphase. We prepared monolayer films of well ordered nanoparticles using the above described technique for various cobalt-platinum nanoparticles of different particle sizes and shapes.

\subsubsection*{Long-range particle ordering}

SEM is not suitable to investigate the surface morphology on the scale of several millimeters in detail. The micrographs show only a small part of the sample. Long-range ordering can be studied by integral diffraction methods. Here GISAXS (grazing incidence small angle x-ray scattering) is the first choice, as it reveals information from a large part (several mm$^{2}$) of the surface. The scattering curves can be compared with the FFT of the SEM images, to demonstrate that the same nanostructure is uniformly spread over a large surface area. From simulations of GISAXS patterns it is possible to get information about the form factor and the interference function, which leads to the shape and size of the particles and their lateral long range order.

Two of the samples, transferred to substrates under 105$^{\circ}$ (cobalt-platinum nanoparticles spherical and cubic, respectively) and one under 150$^{\circ}$ (cobalt-platinum nanoparticles - spherical), were investigated by GISAXS. All GISAXS measurements were performed at the experimental station BW4~\cite{C29} at HASYLAB in Hamburg/Germany (see experimental part). 

The GISAXS patterns, and the curves obtained by slicing the GISAXS patterns along the q$_{y}$ axis at the critical angle of the substrate, as well as the high-resolution SEM images of the samples are presented in the Figure 4. The curves obtained from GISAXS measurements were analyzed using the software "Scatter"~\cite{C30} and the obtained results are given in the Table 1.

\begin{figure}[!h]
\begin{center}
\includegraphics[width=0.45\textwidth]{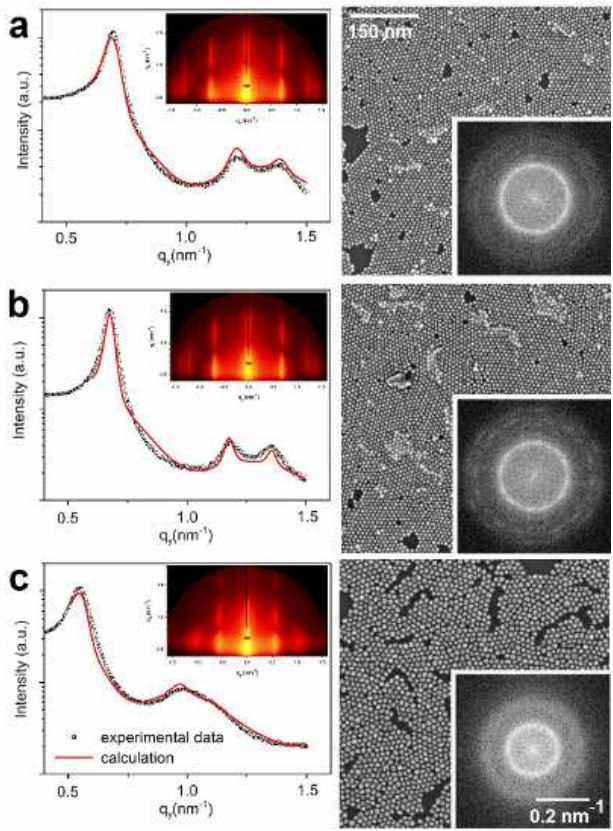}
\caption{\it GISAXS patterns, intensity cuts along q$_{y}$ and fitted curves, and corresponding SEM images of the samples: spherical cobalt platinum nanoparticles deposited under 105$^{\circ}$ (a) and 150$^{\circ}$ (b), and cubic cobalt-platinum nanoparticles deposited under 105$^{\circ}$ (c). FFT of SEM images are shown as insets.}
\label{F04}
\end{center}
\end{figure}

From a qualitative comparison, it is evident that the GISAXS images of the samples prepared from the smaller and spherical particles show a rod-like reciprocal space map corresponding to a very good two-dimensional arrangement of scattering objects. The sample, noted with c, shows broader and less pronounced reflections at higher values of q$_{y}$ indicating lower long-range ordering along the substrate surface. The broad and half-ring-like intensity is the result of scattering from an assembly of objects of similar size (form factor).

\begin{table}[!b]
\begin{center}
\caption {\it Data obtained by GISAXS and FFT analysis of the samples deposited under different angles on silicon substrate.} \label{FE-TABLE} \vspace{0.5cm}
\begin{tabular}{|p{1.2cm}|p{1.2cm}|p{1.2cm}|p{1.2cm}|p{1.2cm}|p{1.2cm}|}
  \hline
  Sample ($\alpha$) & Unit cell (2D-hexa-gonal) [nm] & Relative lattice displacement [\%] & Particle radius (core / core + ligand shell) [nm] & Particle radius standard dev. [\%] & Particle radius (core + ligand shell) FFT [nm] \\ \hline
  \textbf{a} (105$^{\circ}$) & 10.4	& 15.4	& 4.1 / 4.6	& 12.0	& 4.0 \\ \hline
  \textbf{b} (150$^{\circ}$) & 10.7	& 16.8	& 4.2 / 4.6	& 14.0	& 4.0 \\ \hline
  \textbf{c} (105$^{\circ}$) & 13.0	& 23.0	& 5.2 / 6.0	& 12.0	& 5.4 \\ \hline
\end{tabular}
\end{center}
\end{table}

The fitted curves were obtained using the model for hexagonal packed spheres and as can be seen from the graph in Figure 4 they match well with the experimentally obtained curves. The quantitative evaluation of the GISAXS patterns provided a value for the particle radius of about 4.1~nm. The distance of the particles is larger for the sample taken under 150$^{\circ}$ than for the sample taken under 105$^{\circ}$ as the unit cell increases from 10.4~nm to 10.7~nm, respectively (Table 1). This is understood in terms of better particle ordering in sample. For the cubical particles (c) a radius of 5.2~nm (average radius due to approximation of cubic particles with spheres) and the unit cell size of 13.0~nm were calculated from GISAXS data (Table 1). 

FFT analyses of SEM images confirmed the well ordered packing of the particles in monolayers on small areas (approx. 500~nm x 500~nm). The particle radius (core + ligand shell) calculated from FFT analyses is slightly smaller than those calculated from GISAXS measurements (Table 1). These differences, originate most probably from the store resolution (1~pixel = 1~nm) of the SEM images used for the FFT particle size analyses.

The values for the particle radius obtained by GISAXS analysis are in good agreement with the values calculated from both microscopic techniques SEM (Table 1) and TEM (see Supporting Information). The GISAXS analyses confirmed for areas of several mm$^{2}$ the high ordering of the nanoparticle monolayers visible in the SEM images only for a range of a few $\mu$m$^{2}$. Based on the results obtained by GISAXS analyses and SEM images (FFT analyses) we can conclude that lower angles (more horizontal substrates) are better suited for the production of compact films of well ordered nanoparticles. 

\subsubsection*{DC response of nanoparticle films}

Electrical DC measurements were performed on cobalt-platinum nanoparticles with a diameter of 8~nm building monolayer films deposited by the previously described technique. For this purpose silicon wafers with a thermally grown oxide layer of 300~nm thickness were structured with 30~nm thick gold electrodes by e-beam lithography with an interelectrode distance and a width of 1.0~$\mu$m. In a second step the previously presented technique was utilized to deposit cobalt-platinum nanoparticles on the structured surface. Figure 5 shows an SEM micrograph of the electrodes covered with nanoparticles. SEM imaging of the films was performed only after electrical measurements as we observed a considerable change in the electric response of our devices after electron microscopy.

\begin{figure}[!h]
\begin{center}
\includegraphics[width=0.45\textwidth]{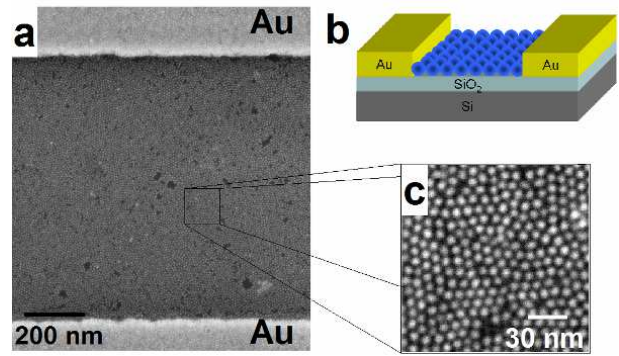}
\caption{\it SEM image of a typical nanoparticle film on e-beam lithographically patterned gold electrodes (a). Sketch of the device (b) and magnified view of the particle film (c).}
\label{F05}
\end{center}
\end{figure}

The room-temperature conductance was extracted by linear fitting of the current-voltage curves yielding 6.5~nS. Figure 6 shows the I-V curves in the range from 80 - 300~K. At low temperatures they show a nonlinear characteristic. This behavior was previously reported for different metal nanoparticle films reflecting the existence of tunnel-barriers in the film resulting in charging energy of single metal particles which determines the electric response of those films~\cite{C06,C03,C31}. An increasing temperature promotes thermal hopping of electrons which overcome the Coulomb blockade. As a consequence the film resistance is rising with lower temperature in contrast to metallic films without tunnel-barriers. The nonlinearity in the DC curves is only observable if the charging energy of particles is higher than the thermal energy. Since our films showed almost ohmic behavior at room temperature further investigations were performed at lower temperatures to explore the charge transport in the deposited films.

\begin{figure}[!h]
\begin{center}
\includegraphics[width=0.45\textwidth]{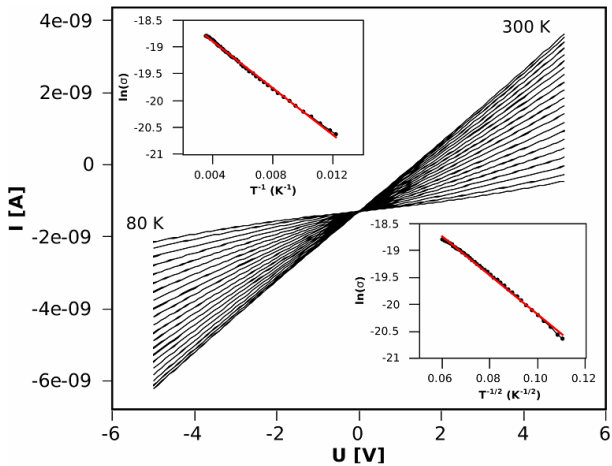}
\caption{\it Current-voltage curves at different temperatures ranging from 80 to 300~K. Upper inset: Arrhenius plot with linear fit, lower inset: plot after the granular metal model with linear fit.}
\label{F06}
\end{center}
\end{figure}

Different transport mechanisms were proposed in the literature. The Neugebauer-Webb model~\cite{C32} describes a simple thermally activated charge hopping between adjacent particles. As a result the conductance $\sigma$ is scaling with

\begin{equation}
	\sigma \propto \exp\left( \frac{-E_{a}}{k T} \right)
\end{equation}
  
where $-E_{a}$ is the activation energy which is equal to the charging energy of particles and $k T$ is the thermal energy. This model fits well to 10.2~nm cobalt-platinum nanoparticle assemblies as we reported previously~\cite{C03}. Linear fit of $\ln(\sigma)$-$T$ plots gave activation energies in the range between 18.5 and 18.9~meV. This value is in good agreement with the results reported by other groups for metal nanoparticles~\cite{C33,C34} and slightly higher compared to the value we obtained previously for 10.2~nm cobalt-platinum nanoparticles ranging from 7-10~meV~\cite{C3}.

In contrast to the Neugebauer-Webb model Abeles et al. suggested an expression which allows calculating the activation energy within the variable range hopping (VRH) formalism proposed for randomly distributed and randomly shaped metal particles in insulating matrices~\cite{C35}. Here the conductance scales with
 
\begin{equation}
	\sigma \propto \exp\left( -2 \sqrt{\frac{-\delta \beta E_{a}}{k T}} \right)
\end{equation}
 
where $\delta$ is the average interparticle distance and $\beta$ is the electron tunneling decay constant. $\beta$ is a function of the tunneling barrier height $\varphi$:
 
\begin{equation}
	\beta = \frac{ \sqrt{2 m_{e} \varphi} }{\hbar}
\end{equation}
 
here $m_{e}$ is the electron mass and $\hbar$ is the Planck constant divided by 2$\pi$, and with the work function of platinum (5.65~eV) $\beta = 1.2 \cdot 10^{10}$~m$^{-1}$. From GISAXS measurements values of 0.7-1.2~nm were calculated for $\delta$. The Figure 7 inset shows the linear fit of the $\ln (\sigma)$-$T^{-1/2}$ data yielding an activation energy of 2.1-2.2~meV. The resulting charging energy is about three times lower than the thermal energy at 80~K ($\sim$7~meV).

Considering the nonlinearity of the current-voltage characteristics at 80~K and close to ohmic behavior at room temperature a value of 7-25~meV (thermal energy at room temperature $\sim$25 meV) can be expected for the charging energy. This suggests that the model of the simple thermal activated charge transport matches better our experimental data.

To better distinguish between the two models the charging energy was additionally calculated from the interparticle capacitance:

\begin{equation}
	E_{a} = \frac{ e_{0}^{2}}{2 C}
\end{equation}
 
This equation explains that the here used 8~nm particles show a larger charging energy compared to the bigger 10.2~nm particles investigated previously~\cite{C03}. The smaller particles possess a lower capacitance yielding higher activation energy. From elementary electrostatics the capacitance of a metallic round core surrounded by an insulating sphere embedded in metallic environment can be derived to be:
 
\begin{equation}
	C = \frac{ 4  \pi \epsilon \epsilon_{0} }{\left( \frac{1}{r} - \frac{1}{r + \delta} \right)}
\end{equation} 
 
The charging energy is then given by equation

\begin{equation}
	E_{a} = \frac{ e_{0}^{2}}{8 \pi \epsilon \epsilon_{0}} \left( \frac{1}{r} - \frac{1}{r + \delta} \right)
\end{equation}  
 
With a value of 2.7 \cite{C36} for $\epsilon$ we attained a charging energy of 13.3~meV which meets the result from the Arrhenius plot (18.5-18.9~meV) much better than the value from the VRH model. This result shows additionally that the simple activated charge transport mechanism is more suitable for our experimental data than the VRH model. 

\section*{Conclusion}

The presented results demonstrate the possibility to deposit ordered films of cobalt-platinum nanoparticles on glycol/air interfaces. These films were successfully transferred onto silicon wafers or samples structured with gold electrodes over unprecedented vast areas in the micrometer up to the millimeter range. High coverage of the substrate and well-ordered packing was obtained with particles of different size, shape, and composition (Co to Pt ratio), as well as under varied substrate angles. 

The film compactness and particle order are influenced by the substrate angle during the film deposition. Nanoparticle films deposited under smaller angle are better ordered, and have the highest coverage. Deposition of the particles onto steeper substrates leads to a good ordering of the nanoparticles but the films undergo breaking during deposition. Additionally, it was shown that ligand excess strongly influences the coverage of the substrate and full coverage can be obtained only by removal of the ligand excess from the nanoparticle solution. 

The charge transport mechanism of monolayers of cobalt-platinum nanoparticles was examined by DC measurements at different temperatures. The simple thermal activated model matched the experimental data better than the VRH approach. The results from DC measurements confirm those from GISAXS and SEM, indicating narrow size distribution of the cobalt-platinum particles and the high order of the achieved nanoparticle films upon vast areas.

\section*{Experimental part}

\textbf{Synthesis:} High quality cobalt-platinum nanocrystals were synthesized via simultaneous reduction of platinum acetylacetonate (Pt(acac)$_{2}$) and thermal decomposition of cobalt carbonyl (Co$_{2}$(CO)$_{8}$) in the presence of 1-adamantan carboxylic acid (ACA) and hexadecylamine (HDA) as stabilizing agents. The used amounts were scaled up twice in comparison to the preparation method of Shevchenko et al.~\cite{C28}, except the amount of Co$_{2}$(CO)$_{8}$, that was slightly increased to 0.092~g (7 \% higher amount than the by original synthesis). Obtained particles were precipitated with 2-propanol, centrifuged and redisolved in chloroform. This procedure was repeated twice in order to remove excess of ligands from the solution (to be publisched). At the end the nanoparticle solution was filtered through a PTFE 0.2~$\mu$m filter. The composition of the obtained particles varies Co$_{14-20}$Pt$_{86-80}$ from the energy dispersive x-ray microanalysis (EDX).

\textbf{Additional washing procedure:} Cobalt platinum nanoparticles were prepared and cleaned as described before. Before the film preparation the moieties of HDA and ACA were washed off in order to remove the excess of ligands. The particles were precipitated by adding three times larger volume of 2-propanol to the stock solution and centrifuged for 10 min. After centrifugation the particles were re-dissolved in chloroform (first additional washing) and the same procedure was repeated once more (second additional washing). 

\textbf{Cobalt-platinum particle films:} were prepared on Langmuir-Blodgett trough NIMA 311D. The program NIMA 516 was used for programming the process and for collecting the data during the production of LB films. For all experiments with water as subphase Millipore water was used. The films were prepared at room temperature (22 $\pm$ 1$^{\circ}$C). 

\textbf{Particle deposition:} The solutions used for preparations of particle films were prepared in the following way: After synthesis the powder of the desired cobalt-platinum nanoparticles was weighted and then dissolved in a known amount of solvent, either toluene or chloroform. The solution was passed through a PTFE filter of 0.45~$\mu$m pore size and stored in clean glassware. 50~$\mu$l glass syringes were utilized to disperse the particle solution uniformly on the water/glycol interface. The solvent evaporated usually in 10-15~min after deposition. Pressure/area isotherms on the water/air interface were recorded using a surface compression rate of 30~mm/min until the surface pressure started to increase and then continued with 5~mm/min until the end of the measurement.

For the nanoparticle films deposited on glycol/air interface, the isotherms were obtained in the same way. At the point where the surface pressure starts to increase (defined on the basis of the isotherms obtained as described) the pressure program was applied referred to as "Isothermal cycle". Each compression cycle consisted of a particle compression in steps of 5~cm$^{2}$ and relaxation time of 10~s till next compression step. Altogether 100 cycles were applied. Near the target pressure (8~mN/m) the barrier movement was limited by the program between 8.5 and 7.5~mN/m. All films were deposited onto cleaned silicon wafers (8x8~mm). The dipper speed was 1~mm/min. Obtained films were further dried for 24~h in a vacuum oven and kept afterwards at ambient conditions.

\textbf{Scanning Electron Microscope (SEM):} images were recorded using LEO 1550 scanning electron microscope (spatial resolution of $\sim$1~nm). 

\textbf{GISAXS measurements} were performed using the grazing incidence setup of the experimental station BW4~\cite{C29} at HASYLAB (Hamburg/Germany) equipped with a high resolution 2-dimenional CCD detector (MAR research, 2048x2048 pixel, pixel size 79~$\mu$m) at a distance of 2.5~m (sample-detector). The wiggler beam line was set to a wavelength of 0.138~nm. The flight path was fully evacuated and the beam size was focused by an additional beryllium lens system to a size of only 30~$\mu$m (vertical) x 60~$\mu$m (horizontal) at the sample position. The incident angle of the primary beam was 0.52$^{\circ}$. Piezo driven slits were installed in front of the sample to reduce diffuse scattering from the collimation devices of the beam line. With this setup it was possible to determine structures within a scale of some nanometers up to around 400~nm.

\textbf{DC measurements:} The electrical measurements were performed with the Agilent 4156C Precision Semiconductor Parameter Analyzer. The measurements at room temperature were performed on a home-build probe station. The wires were double electrostatic shielded as far as possible. For low temperature measurements an Oxford Instruments OptistatCF cryostat was used.

\clearpage

\end{document}